\documentclass[preprint,12pt]{elsarticle}

\usepackage{amsmath}
\usepackage{mathtools}
\usepackage{amsfonts}
\usepackage{amssymb}
\usepackage{graphicx}

\biboptions{sort&compress}

\journal{Current Opinion of Structural Biology}

\begin{document}

\begin{frontmatter}



\title{Structural ensembles of disordered proteins from hierarchical chain growth and simulation}


\author[a]{Lisa M. Pietrek\fnref{contrib}}

\author[b,c,d]{Lukas S. Stelzl\fnref{contrib}}

\author[a,e]{Gerhard Hummer\corref{cor1}}
\ead{gerhard.hummer@biophys.mpg.de}
\address[a]{Department of Theoretical Biophysics, Max Planck Institute of
  Biophysics, Max-von-Laue-Stra{\ss}e 3, 60438 Frankfurt am Main, Germany}
\address[b]{
Faculty of Biology, Johannes Gutenberg University Mainz, Gresemundweg 2,
55128 Mainz, Germany}
\address[c]{KOMET 1, Institute of Physics, Johannes Gutenberg
University Mainz, 55099 Mainz, Germany}
\address[d]{Institute of Molecular
  Biology (IMB), 55128 Mainz, Germany}
\address[e]{Institute for Biophysics, Goethe University, 60438 Frankfurt am
    Main, Germany}

  \fntext[contrib]{Authors contributed equally}
  \cortext[cor1]{Corresponding author}
  

\begin{abstract}
Disordered proteins and nucleic acids play key roles in cellular function and disease. Here we review recent advances in the computational exploration of the conformational dynamics of flexible biomolecules. We focus on hierarchical chain growth (HCG) from fragment libraries built with atomistic molecular dynamics simulations. HCG combines chain fragments in a statistically reproducible manner into ensembles of full-length atomically detailed biomolecular structures. The input fragment structures are typically collected from molecular dynamics simulations, but could also come from structural databases. Experimental data can be integrated during and after chain assembly. Applications to the neurodegeneration-linked proteins $\alpha$-synuclein, tau, and TDP-43, including as condensate, illustrate the use of HCG. We conclude by highlighting the emerging connections to AI-based structural modeling.
\end{abstract}


\begin{highlights}
\item Hierarchical chain growth (HCG) produces extensive ensembles of disordered proteins with atomic detail.
\item HCG ensembles capture key experimental descriptors “out of the box” and can be refined against experimental data where needed.
\item HCG builds on successful data-driven statistical coil models in structural biology and polymer theory, and complements molecular dynamics (MD) simulations.
\item In applications to autophagy, neurodegeneration-linked proteins $\alpha$-synuclein, tau, and TDP-43, including as phase-separated condensate, HCG has revealed how local structural characteristics affect biological (dys)function.
\end{highlights}


\end{frontmatter}

\section*{Introduction}

A significant fraction of the proteome in higher organisms consists of intrinsically disordered proteins (IDPs) that do not fold into well-defined structures and of proteins with intrinsically disordered regions (IDRs) \cite{Wright1999}. Disordered segments are also present in nucleic acids. In particular, single-stranded RNAs (ssRNAs) such as messenger RNA (mRNA) feature regions that do not form double helices or other folded structures \cite{Grotz2018,Bottaro2018}. 
IDPs and IDRs are unfolded in solution and can transiently adopt secondary structure \cite{Mukrasch2009}. Binding to other biomolecules can induce IDRs to fold \cite{Ulmer2005}, though disorder can persist also in the bound state 
\cite{Borgia2018}. IDPs and IDRs have distinct functions, e.g., in the nuclear pore complex \cite{Yu2022}, are a major component of biomolecular condensates \cite{Li2012}, and are closely linked to neurodegenerative diseases 
\cite{Chen2019} with their interactions (dys)regulated by mutations and post-translational modifications \cite{Ambadipudi2017,Wegmann2018}.

The structural heterogeneity of IDPs is best represented by a broad structural ensemble \cite{Bernado2005fm}. 
Non-local interactions in IDPs are necessarily transient, unlike in folded proteins. As a consequence, the conformation space of IDPs is inherently hierarchical in the sense that, at any scale, the local conformational preference will be minimally impacted by regions distant in sequence. Building on this principle, we recently introduced hierarchical chain growth (HCG) \cite{Pietrek2019}  
to explore the structural heterogeneity of IDPs.

Here, we review the concepts and applications of HCG by computational fragment assembly as an extension, alternative, and complement to molecular dynamics (MD) simulations for IDPs. By preserving the local structure across scales where possible, chain growth is appealing not only because of high computational speed and flexibility, but also by the possibility to produce accurate representations of the structural ensembles even of large IDPs. Chain growth can be used to create a broad ensemble of structures that can, if needed, be refined by integrative modeling using experimental data and/or MD simulations.

\begin{figure*}[tbhp]
\centering
\includegraphics[width=\textwidth]{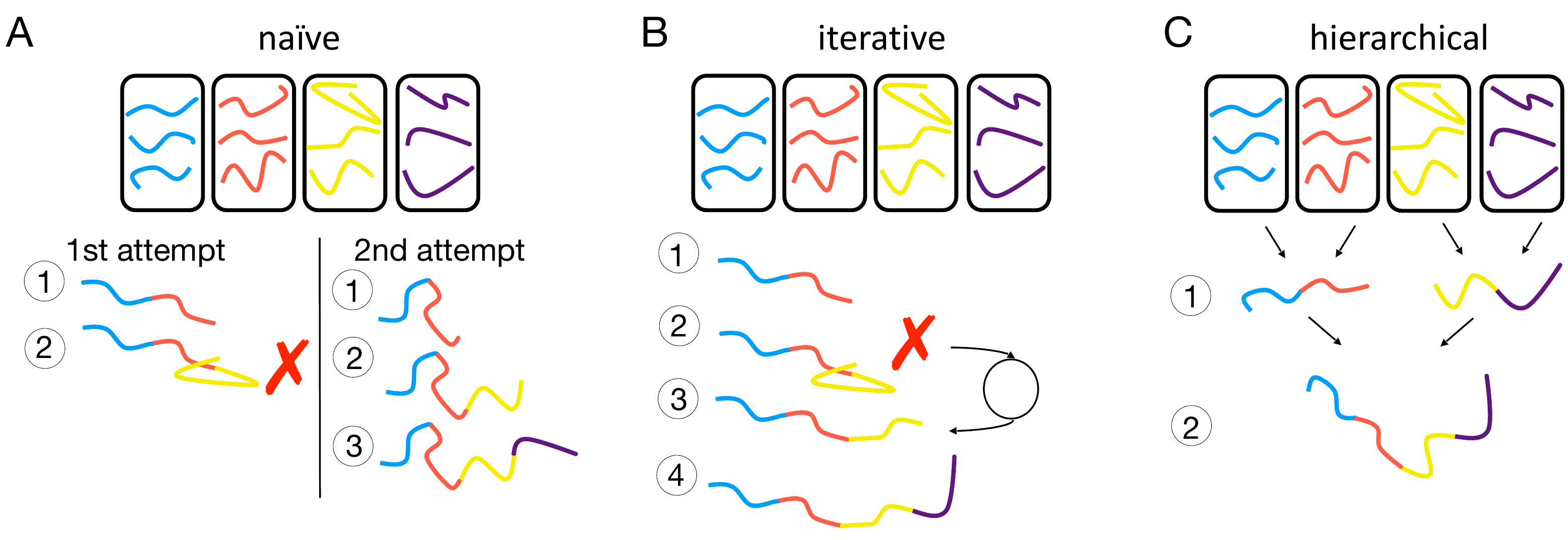}
\caption{\textbf{Schematic of naive, iterative, and hierarchical chain growth}. The structures of a linear biopolymer are assembled from four fragments (colored chains) picked from their respective pools (ovals). (A) Growing chains by a naive algorithm. On encountering a clash, the current chain is rejected and a fresh attempt is launched. While correct, this algorithm is extremely inefficient for long chains. (B) Iterative algorithm. Instead of re-growing the entire chain when a clash is detected, many chain-growth approaches simply repeat the step until a conformation without clash is obtained. Such algorithms are incorrect unless the bias resulting from repeated drawings is properly accounted for, as in Rosenbluth sampling. (C) Hierarchical chain growth (HCG) is a correct and efficient algorithm. Different fragments are recursively combined until the full-length chain is obtained. Absent steric clashes, monomer fragments are combined to dimers, dimers to tetramers and so on. For chains with $N=2^M$-fragments, the algorithm has only $M=\log_2 N$ assembly levels.} 
\label{fig:concepts}
\end{figure*}

\section*{Chain growth }

Modeling of the global structure of polymers has long been approached by chain growth algorithms. For a biomolecule with internal structure, we imagine dividing its sequence into fragments (Figure~\ref{fig:concepts}). For each of these fragments, we generate a pool of structures, as illustrated schematically with the four urns in Figure~\ref{fig:concepts}. This pool may be filled with local structures taken from databases of experimental structures or from molecular dynamics simulations of chain fragments. The task is then to assemble these fragments by a chain-growth algorithm. Naively one might consider that one simply needs to grow polymer chains sequentially, say from N to C terminus (Figure~\ref{fig:concepts}A). However, so not to introduce a bias, one would have to stop the growth of a chain as soon as a clash is encountered and start to grow an entirely new chain instead of simply redrawing a new fragment (Figure~\ref{fig:concepts}B). Otherwise, the outcome will depend on arbitrary choices such as the direction of chain growth, N-to-C versus C-to-N.
Rosenbluth and Rosenbluth recognized this problem of detailed balance in chain growth early in the history of computer simulations, and addressed it by a careful reweighting of self-avoiding random walks (SAWs) on a lattice \cite{Rosenbluth1955a}.

In combination with importance sampling, chain growth has become a powerful tool to create large ensembles for polymers, including biopolymers \cite{IljaSiepmann1992,Lettieri2011}.  
To grow a chain, one assembles short fragments that can be sampled very efficiently at high quality. For IDPs, the flexible-meccano model by Bernad\'{o} et al. \cite{Bernado2005fm} 
is widely used, also for proteins under physiological conditions \cite{Adamski:JACS:2019}. It builds on the observation that the local structure in IDPs is captured well by coil models \cite{schwalbe1997structural,Feldman2000a,Dobson2001,Fitzkee2004,Krzeminski2013,Estana2019}. In flexible-meccano, chains are grown based on the backbone-dihedral statistics in the Protein Data Bank (PDB).

\section*{Hierarchical chain growth}

In disordered proteins, local structure is determined primarily by the local amino acid sequence, lacking the cooperative interactions of folded proteins between regions distant in sequence. 
HCG \cite{Pietrek2019} exploits this hierarchical nature. A protein chain is divided into overlapping sequence fragments. Fragment structures are sampled with replica-exchange molecular dynamics (REMD) simulations. From the resulting pools, the fragments are then chosen at random. Adjacent fragments are combined with a rigid body superimposition of the heavy atoms of their overlapping regions. If the corresponding root-mean-square distance (RMSD) is below a given cut-off and if there are no steric clashes, the fragment pair is entered into the respective pool at the next assembly level. This assembly process is continued hierarchically all the way up to the level of full-length chains (Figure~\ref{fig:concepts}C). 
At each level of the assembly process, the size of the chain fragments effectively doubles.
The hierarchical assembly manifestly preserves detailed balance,
which guarantees that arbitrary choices such as the order of the assembly do not affect the final ensemble. Hence, HCG grows ensembles of chains with a well-defined distribution. By construction, the members of the HCG ensemble are statistically independent. As a result, HCG produces broad ensembles of IDPs with highly diverse conformations in a computationally efficient manner, sampling a significantly broader conformational space than, say, one 2 $\mu$s-long MD simulation in case of $\alpha$-synuclein (aS) \cite{Pietrek2019}. 

\begin{figure*}
\centering
\includegraphics[width=\textwidth]{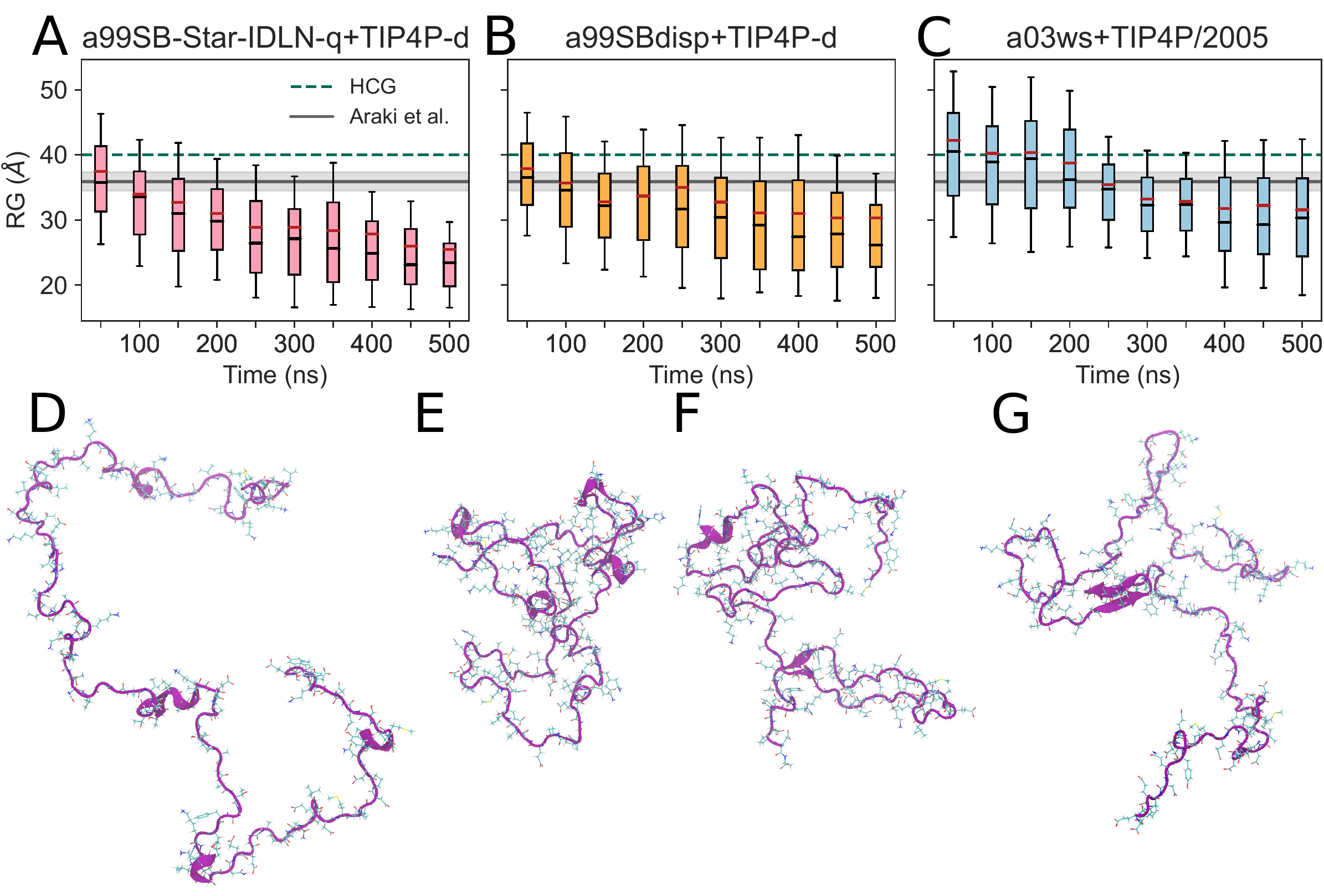}
\caption{HCG of $\alpha$-synuclein extended by atomistic MD simulations with different force fields. (A-C) The box-and-whiskers plots show the distribution of the radius of gyration R$_G$ calculated over windows of 50 ns across 20 independent runs initiated from 20 randomly chosen structures of the HCG ensemble (mean: black; median: red; box: interquartile range; bars: extrema). Results are for (A) the amber99StarILDN-q force field and TIP4P-d water model \cite{Piana2015}, (B) the a99SBdisp force field and TIP4P-d water model \cite{Robustelli2018}, and (C) the a033ws force field and TIP4P/2005 water model \cite{Best2014b}. The dashed green line indicates the average RMS R$_G$ for a 20000 HCG ensemble of aS. The solid gray line is the R$_G$ value measured via SEC-SAXS by Araki et al. \cite{Araki2016a} with the standard error indicated by shading. (D-G) Snapshots of aS as grown with HCG before MD (D), and after 500 ns MD (E-G) with the force fields of panels A-C.}
\label{fig:MD_HCG_sampling_aS}
\end{figure*}

If needed, HCG can be complemented by MD simulations of solvated full-length chains. As shown for aS in Figure~\ref{fig:MD_HCG_sampling_aS}, the radius of gyration $R_G$ calculated for an HCG ensemble with 20,000 chains \cite{Pietrek2019} is already in good agreement with the measured value from SEC-SAXS \cite{Araki2016a}. For three different combinations of protein force field and water models, we found that aS tended to collapse below the size seen in the SEC-SAXS measurements \cite{Araki2016a}. These findings highlight, first, that care must be taken to assess the collapse tendency. Second, as shown in Figure~\ref{fig:MD_HCG_sampling_aS}, even for the loosely packed aS with 140 amino acids, it takes many hundreds of nanoseconds of MD just to relax the chain size. Third, without any further simulations, HCG appears to be at least on par with the three MD simulation models. HCG thus provides an excellent starting point for further inquiry.

Applications of HCG extend beyond the sampling of IDP ensembles. For instance, HCG has shed light on the early stages of autophagy. Sawa-Makarska et al. used an implementation of HCG to model the disordered N and C termini of the protein Atg9 in the Atg9-containing vesicles seeding yeast autophagosomes \cite{Sawa-Makarska2020}. The extensive coverage of the vesicle surface by the Atg9 tails explained the relatively low rate of Atg8 lipidation, which requires unhindered access to the surface. Interestingly, some of the principles used in chain growth also find their application in other approaches to model important biological systems such as glycoproteins. Gecht and co-workers implemented a tool, GlycoSHIELD \cite{Gecht2022}, that helped, e.g., to prepare a proper model of the SARS-CoV-2 spike protein by attaching glycan conformers onto the protein of interest. In another variant, Turo{\v{n}}ov{\'{a}} et al. \cite{Turonova2020} resampled the hip and knee joints of SARS-CoV-2 spike stalk to probe the full extent of its mobility.

Interactions between distant parts of the chain other than steric exclusion can be taken into account \cite{Lettieri2011,Stelzl2022}, including electrostatics, at least at the level of implicit solvent descriptions. Including electrostatic forces in HCG may be important for growing structures of highly charged biomolecules \cite{Borgia2018}.
A pragmatic way forward can be to use larger chain fragments for HCG sampled in MD simulations using explicit ionic solvents.


\begin{figure*}
\centering
\includegraphics[width=\textwidth]{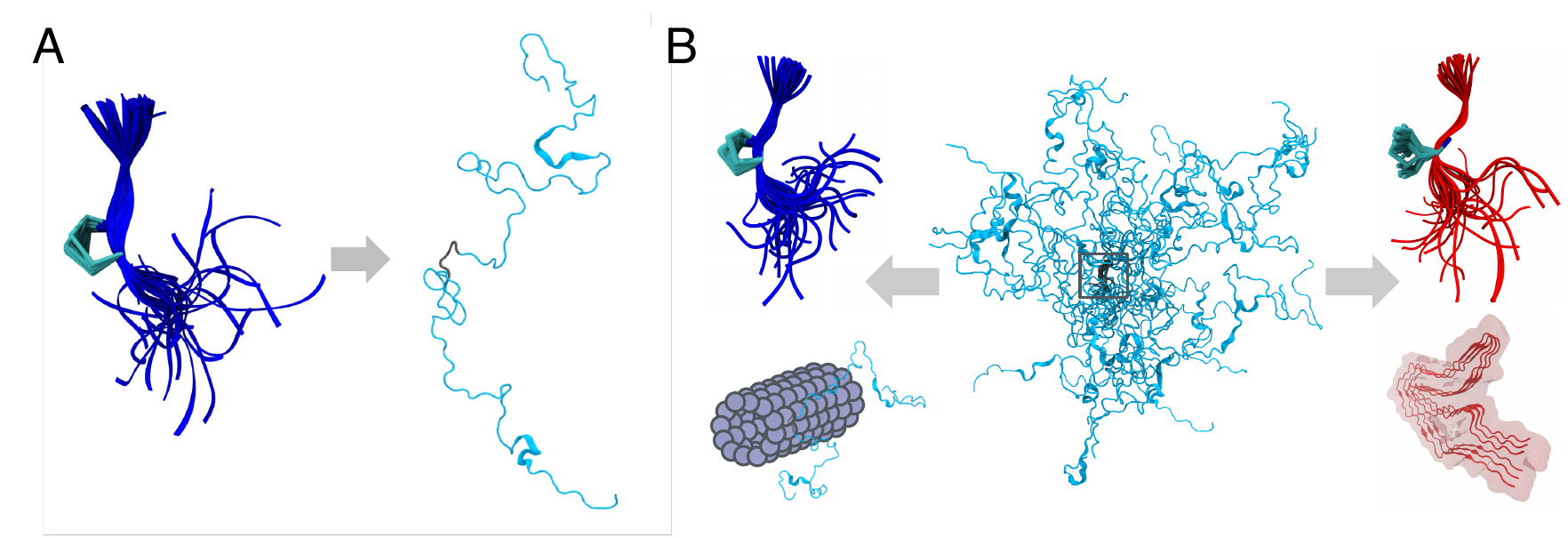}
\caption{\textbf{Chain growth captures shifts from functional to aggregation-prone structural forms in tau K18.} (A) Local structures overlaid at residues V300-G304, with P301 shown as licorice. This segment is indicated in gray in full-length K18. (B) Changes in the conformational preference of the V300-G304 region contribute to a shift from functional binding to microtubuli (left) dominant in WT (blue) to fibril formation amplified by the P301L mutation (red). Data and figures adapted from Ref. \cite{Stelzl2022}, which was published under Creative Commons BY 4.0 license.}
\label{fig:overview_tau_aS}
\end{figure*}

\section*{Integration of experimental data}

An ensemble representation establishes a sound foundation for the interpretation of experimental data in case of structural disorder in a molecular system \cite{Reichel2018,Kofinger2019,Larsen2020,Bottaro2018,Tesei2020}.
As a first line of attack to improve the consistency between measured and calculated observables, one can reweight the members of the unbiased ensemble rather than adjust their structure \cite{Rozycki2011,Boomsma2014a,Hummer2015,Bottaro2020}. In a Bayesian view, the initial ensemble can be considered a sample of the prior distribution. By imposing restraints derived from experiment already in the creation of the ensemble \cite{Lindorff-Larsen2005,Papaleo2018,Hummer2015}, this sample can be enriched. Combinations with enhanced sampling techniques such as metadynamics \cite{Bonomi2016} or replica exchange \cite{Hummer2015} further improve the sampling efficiency. Uncertainties in measurements and their modeling are readily taken care of in a Bayesian framework \cite{Hummer2015}. However, the integration of data is no panacea: for comparably poor force fields, the overlap with the ``true'' ensemble may not be sufficient for reweighting according to a single or a few experimental observables to establish meaningful ensembles \cite{Ahmed2021}. In other words, the quality of the Bayesian prior matters, which may not surprise considering the vast conformational space to be sampled.

In chain growth, experimental data can be integrated already during the ensemble generation in a form of integrative modeling \cite{Krzeminski2013}. 
The flexible-meccano approach and its extension ASTEROIDS have been successfully used to account for different types of NMR data and single-molecule FRET and SAXS data \cite{Naudi-Fabra:JACS:2021}. 
Biased fragment choice, with fragment weights derived from a Bayesian formulation, has been shown to be powerful in early applications of chain growth \cite{fisher2010modeling} 
or in the refinement of MD ensembles of flexible proteins by fragment replacement \cite{Boomsma:14}.

The reweighted hierarchical chain growth (RHCG) is an extension of HCG to integrate experimental data by assigning weights to the fragment conformations \cite{Stelzl2022}.
RHCG is designed to counteract the problem of systematic biases in the fragment pool. Consider, for instance, a systematic force-field error in the energetic balance between locally extended and helical peptide conformers. As the size of the molecules increases, it becomes less likely that all parts of a chain are drawn from the relevant subspace. Consequently, after global reweighting only a few chains may end up dominating the final ensemble. RHCG counteracts this tendency by using suitable fragment weights, which can be assigned, for instance, by Bayesian inference \cite{Hummer2015,Kofinger2019,Bottaro2020}. In a global reweighting of the ensemble after chain assembly, the fragment weights are fully accounted for \cite{Stelzl2022}. In this way, RHCG generates a well-defined and diverse output ensemble that has high overlap with the true ensemble.

With RHCG we were able to account for solution experiments on tau K18 as diverse as NMR, single-molecule FRET, and small-angle X-ray scattering \cite{Stelzl2022}. We also captured structural features seen in tau fibrils and provided important atomic-resolution insight to complement current ideas of how tau mutations shift the balance between the tau conformational ensembles in health and disease (Figure ~\ref{fig:overview_tau_aS}B). P301L, P301T, and P301S mutations shift our tau ensembles away from turn-like conformations that would be able to bind to microtubules. Instead, this region populates aggregation-prone extended conformations. RHCG has thus revealed how subtle shifts in local-structural propensities could give rise to pathogenesis \cite{Stelzl2022}. 

Teixeira et al. \cite{teixeira2022} recently published a software suite that samples IDP ensembles following the principles of data-driven coil models and contains tools for further analysis and ensemble refinement.
Interestingly, their approach also captured shifts in local structures propensities in response to the neurodegeneration-linked P301L mutations in accordance with the RHCG ensemble \cite{Stelzl2022}. 

\section*{Condensates}

\begin{figure}
\centering
\includegraphics[width=0.5\textwidth]{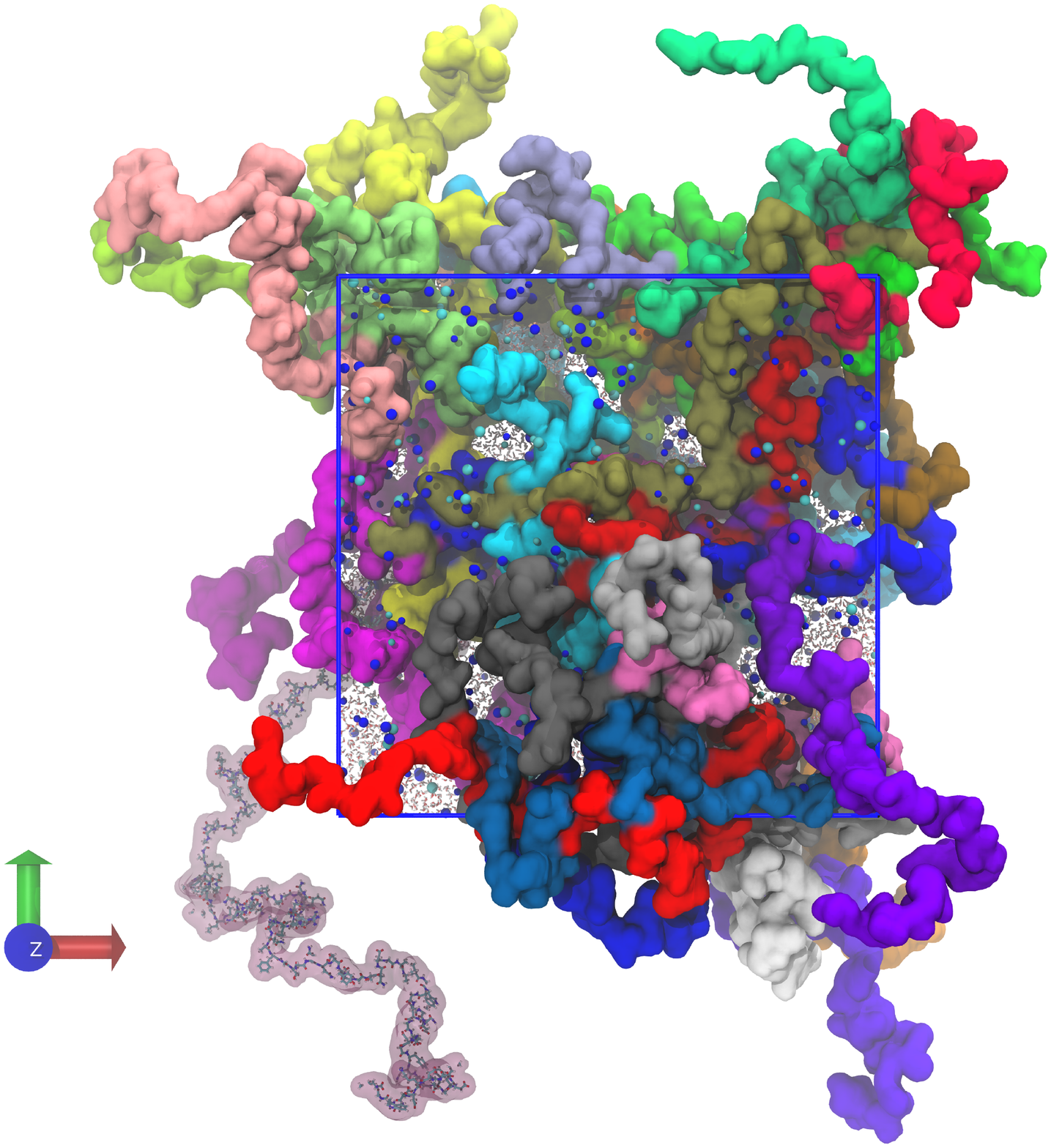}
\caption{\textbf{Snapshot of an LCD TDP-43 condensate.} The all-atom system was built for MD simulations by combining TDP-43 LCD chains preassembled by HCG \cite{GruijsdaSilva2022}. The surface of the chains is shown in color and atoms from a single TDP-43 LCD chain are shown with atomistic detail. Solvent molecules are omitted for clarity except for a small region, where water is shown as sticks and ions as spheres (sodium in cyan, chloride in blue). Blue lines indicated the periodic simulation box.}
\label{fig:dense_solution}
\end{figure}

IDPs are often associated with protein condensates. An exciting perspective is to build molecularly detailed models of such crowded solutions of (disordered) biomolecules. One possibility is to harness the power of HCG to directly model such dense systems. Individual conformations are drawn from an ensemble of single chains grown with HCG and assembled in a simulation box, which can serve as starting point for MD simulations. For the low-complexity domain (LCD) of the neurodegeneration-linked RNA-binding protein TDP-43, we generated models of condensates  with atomic detail (Figure \ref{fig:dense_solution}) using a variant of HCG and then ran MD simulations from this initial system \cite{GruijsdaSilva2022}. 
In the simulations,  phosphomimicking mutations led to a loss of protein-protein interactions and an increase in protein solvent interactions in the C terminus of the TDP-43 LCD that destabilized the condensates, complementing coarse-grained simulations of the phase behaviour of phosphomimicking mutants and phosphorylated TDP-43. The experiments 
by Dormann and colleagues \cite{GruijsdaSilva2022} have suggested that disease-linked phosphorylation, rather than driving the progression of neurodegenerative diseases, is a potential cell-protective mechanism; by hyperphosphorylation the cell may try to hinder the condensation and aggregation of TDP-43.

Combining high-resolution experiments, theory, atomistic and coarse-grained modeling has already started to yield insights into the drivers of liquid-liquid phase separation \cite{Martin2020}. This is a particularly exciting prospect as coarse-grained simulation models parameterized using large sets of high-resolution experimental data can capture trends in the global arrangements of disordered proteins as well as their propensities to phase separate \cite{tesei:pnas:2021}.
Another interesting direction is the simulation of dense solutions of disordered proteins or their fragments at sub-critical concentrations \cite{Paloni2020}. 
Such simulations \cite{Paloni2020,Polyansky2022,Murthy2019} can provide critical insights into molecular driving forces for condensation. However, how to optimally combine models from chain-growth and atomistic simulations with coarse-grained models is an open question.

\section*{Outlook}

On the methods side, the emerging connections of chain growth to machine learning and artificial intelligence (AI) deserve special attention. Historically, coil models have been an attempt to collect statistical information about protein structure and use it to infer the local and global structures of proteins. As such, coil models and HCG have a natural connection to machine learning and AI.

AlphaFold2 \cite{Jumper2021} showcases the power of AI to predict three dimensional protein structure. The resulting acceleration in structural studies of complex assemblies \cite{Mosalaganti2022} raises the intriguing question as to what can be learned about disordered regions from AlphaFold predictions. Currently AlphaFold2 does not capture disordered regions as a properly weighted ensemble. Hence, an exciting prospect is the combination of AlphaFold2 models of the folded protein and conformations from IDP/IDR ensembles using, e.g., HCG, molecular dynamics or knowledge-based approaches. Interestingly, segments in IDRs often appear structured in AlphaFold2 predictions, possibly in reflection of their binding to distinct partner proteins \cite{Alderson2022.02.18.481080}, which had been used effectively to map and model the interactions of short linear motifs (SLiMs) with structured nucleoporins in the scaffold of the nuclear pore complex \cite{Mosalaganti2022}. 
One potential problem is that AlphaFold2 may capture, in the same model, structures an IDR may adopt in different complexes, as has been shown for conditionally folded proteins by comparison to experimental structures. Thus, a critical assessment of the thousands of local structures predicted for IDPs/IDRs may be advisable even for proteins where AlphaFold2 produces high-quality models of the folded domains. 

Growing efforts have also been made to harness the power of AI to characterize structural ensembles of IDPs. Gupta and colleagues recently developed an AI based approach that learns IDP conformational space from short MD simulations to then generate broad IDP ensembles \cite{Gupta2022}. It is interesting to speculate to what extent this approach can be combined with ensembles sampled with HCG. Zhang et al. \cite{zhang2022} 
are developing a neural network that learns structural ensembles of disordered proteins from experimental information. In fact, the neural network generates and learns torsion-angle probability distributions for interdependent neighboring residues, while also biasing the probability distribution towards experimental data, using a Bayesian formalism. Even more ambitiously, a recent preprint shows how a coarse-grained representation of an atomistic ensemble can be learned by a neural network, which reproduces the equilibrium densities of the input ensemble \cite{Koehler2022}.
HCG ensembles usually extend beyond the conformations sampled by direct MD simulations, at least for long chains, and should thus provide a valuable reference in these endeavors. 

In recent years, we have witnessed a lot of progress in sampling structural ensembles of flexible (bio)polymers. However, efficient sampling of the vast conformational diversity still remains challenging. Approaches that model conformational ensembles based on local structure statistics, i.e., coil models, have been shown to be promising. The hierarchical chain growth (HCG) builds on the basic ideas of coil models. Using HCG one can sample ensembles with highly diverse conformations in a computationally efficient manner. In the cases studied, the ensemble properties agreed well with available experimental data. The quality of the generated ensemble could be further improved by integrating experimental information, producing richly detailed structural ensembles consistent with experiment across scales.


\section*{Acknowledgments}

We acknowledge financial support from the German Research Foundation (CRC902: Molecular Principles of RNA Based Regulation) and the Max Planck Society. L.S.S  thanks ReALity (Resilience, Adaptation and Longevity), M$\mathrm{^3}$ODEL (Mainz Institute of Multiscale Modeling) and Forschungsinitiative des Landes Rheinland-Pfalz for their support. Funded by the Deutsche Forschungsgemeinschaft (DFG, German Research Foundation) - Project number 233630050 - TRR 146.

We thank J\"{u}rgen K\"{o}finger, Iva Pritis\u anac, Kresten Lindorff-Larsen, Matteo Paloni, Matthieu Chavent, Dorothee Dormann, Ben Schuler and Markus Zweckstetter for many insightful discussions.

\bibliographystyle{currbiolX}
\bibliography{cosbCompact}











\end{document}